\documentclass[a4paper,10pt]{IEEEtran}
\renewcommand{\baselinestretch}{0.99}
\usepackage{cite}
\usepackage{amsmath,amsthm,amssymb,amsfonts,bm,mathtools}
\usepackage{graphicx,flushend}
\usepackage{algorithm,algorithmic}
\usepackage{subcaption}
\usepackage{xcolor}

\usepackage{verbatim}

\def\BibTeX{{\rm B\kern-.05em{\sc i\kern-.025em b}\kern-.08em
    T\kern-.1667em\lower.7ex\hbox{E}\kern-.125emX}}

\newtheorem{theorem}{Theorem}[section]

\newtheorem{proposition}[theorem]{Proposition}

\newtheorem*{remark}{Remark}

\newcommand{\h}{\mathbf{h}}

\newcommand{\n}{\mathbf{n}}

\newcommand{\p}{\mathbf{p}}

\renewcommand{\v}{\mathbf{v}}

\newcommand{\y}{\mathbf{y}}

\newcommand{\J}{\mathbf{J}}

\newcommand{\R}{\mathbf{R}}

\newcommand{\U}{\mathbf{U}}

\newcommand{\Compl}{\mbox{$\mathbb{C}$}}

\renewcommand{\Im}{\mathrm{Im}}

\renewcommand{\Re}{\mathrm{Re}}

\begin{document}

\title{Simultaneous Secure Communications and Sensing Privacy with Full Duplex MIMO Radios}

\author{\IEEEauthorblockN{
Ioannis Gavras$^1$ and George C. Alexandropoulos$^{1,2}$ 
} 
\\
\IEEEauthorblockA{$^1$Department of Informatics and Telecommunications, National and Kapodistrian University of Athens, Greece}
\IEEEauthorblockA{$^2$Department of Electrical and Computer Engineering, University of Illinois Chicago, USA}
\\
\IEEEauthorblockA{e-mails: \{giannisgav, alexandg\}@di.uoa.gr}\vspace{-0.5cm}
\thanks{This work has been supported by the SNS JU project 6G-DISAC under the EU's Horizon Europe research and innovation programme under Grant Agreement number 101139130.
}}
\maketitle
\thispagestyle{empty}
\begin{abstract} This paper proposes a Full Duplex (FD) framework for safeguarding Integrated Sensing and Communications (ISAC) against a passive multi-antenna Eavesdropper (Eve). The adversary is modeled as both a communication and sensing threat, seeking to decode uplink data streams and estimate the surrounding scene. Eve's threat is quantified through two worst-case requirements over a location uncertainty region: a minimum secrecy rate for data confidentiality and a minimum eavesdropping Cram\'er-Rao Bound (CRB) for sensing privacy. We derive the eavesdropping information measures governing both data interception and scene inference, showcasing that transmitting Artificial Noise (AN) can degrade Eve's capabilities through the inverse interference-plus-noise covariance. Exploiting the underlying structure of the problem, we characterize finite-power local and robust feasibility conditions for target secrecy-privacy pairs. The analysis showcases that AN is fundamentally subspace-limited: increasing power is beneficial only along Eve's communication channel and sensing-derivative directions illuminated by the effective AN subspace. Finally, we formulate an AN power allocation problem that is solved via monotone branch-and-bound, and numerical evaluations quantify the AN power required to guarantee joint secrecy and sensing privacy.
\end{abstract}


\section{Introduction}
The Integrated Sensing and Communications (ISAC) paradigm is transforming wireless networks from pure connectivity platforms into systems capable of jointly delivering communication and sensing services, including localization, tracking, and environmental awareness~\cite{6G-DISAC_mag,stylianopoulos2025distributed_disac}. 
Although this dual objective improves spectral and hardware usage efficiency \cite{su2026integrating_secure_isac,rexhepi2025blinding_ris}, it also creates a distinct security and privacy challenge: the same waveform that carries the communication payload also reveals information about the surrounding physical scene~\cite{Gavras2026RIS,han2026sensing_secure_isac}. To this, an unintended Receiver (RX) may exploit the geometric structure of the received signal to infer the surrounding scene, even when data payload is protected by conventional means. Secure ISAC must therefore address both message confidentiality and sensing privacy~\cite{qu2024privacy_security_isac}, i.e., the implicit leakage of environmental information through the wireless channel \cite{su2026integrating_secure_isac}.

The ISAC open technical literature has extensively studied waveform design, beamforming, and communication--sensing tradeoffs~\cite{liu2021cramer,xiong2023fundamental}, with recent works further incorporating reconfigurable metasurfaces~\cite{RIS_ISAC}, presenting distributed architectures~\cite{6G-DISAC_arch}, and taming the near-field regime~\cite{zhao2024modeling}. In parallel, physical-layer security techniques, including Artificial Noise (AN) injection~\cite{AN_BEP}, secrecy rate optimization, and spoofing, have been developed to counter eavesdropping~\cite{Gavras2026RIS,han2026sensing_secure_isac,10143983}. Yet, most such formulations focus on message interception, whereas sensing-privacy leakage in ISAC remains less understood~\cite{qu2024privacy_security_isac}. To this end, a suitable safeguarding criterion would account for both Eavesdropper's (Eve) decoding capability and their ability to extract geometric information from the received waveform. This calls for a joint secrecy-sensing-theoretic treatment, where communication is measured through the secrecy rate~\cite{11205098}, while sensing leakage is captured through Fisher-information-related metrics. Full Duplex (FD) radios~\cite{9933358,Islam_2022_ISAC,smida2024inband_full_duplex_isac} are well suited to this setting as they enable simultaneous uplink reception and downlink AN transmission over the same time-frequency resources~\cite{gavras2024simultaneous,lei2021fd_security_an}; however, their use as an active countermeasure for jointly protecting communication secrecy and sensing privacy remains largely uncharacterized.


Motivated by this gap, in this paper, we study the potential of FD Multiple-Input Multiple-Output (MIMO) systems~\cite{9933358} to safeguard wideband ISAC under a joint communications and sensing threat. Eve is modeled as an adversary that must be simultaneously constrained in terms of its achievable information rate and its sensing accuracy over a location uncertainty region. In the uplink, the FD node receives the User Equipment's (UE) Orthogonal Frequency-Division Multiplexing (OFDM) waveform for data decoding and scene inference, while, in the downlink, it transmits AN to impair Eve's interception capability. The proposed formulation couples a worst-case secrecy rate requirement with a sensing-privacy requirement based on the Cram\'er-Rao Bound (CRB), and exposes the finite-power limits of AN-aided protection. In particular, we showcase that AN's effectiveness is governed by inverse-covariance weighting and is fundamentally subspace-limited. Exploiting the underlying structure of the problem, we derive feasibility conditions for secrecy-privacy target pairs, formulate a branch-and-bound AN power allocation method that certifies the minimum required power, and numerically quantify the AN power needed to guarantee satisfactory joint secrecy and sensing privacy performance.

\section{System and Signal Models}
\begin{figure}[!t]
	\begin{center}
	\includegraphics[width=0.9\columnwidth]{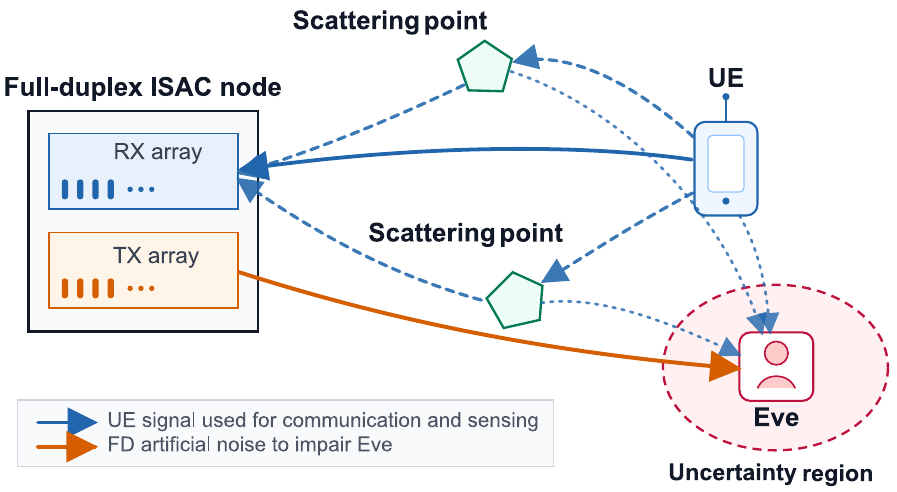}
	\caption{\small{System model of the considered FD-enabled ISAC wireless setup, illustrated for the representative case of two scattering points. The FD node receives the UE uplink waveform through its RX array and exploits the direct and single-bounce reflected paths for data decoding and scene sensing. At the same time, its Transmit (TX) array emits artificial noise toward Eve to degrade both communication interception and sensing inference capabilities.
   }}
    \vspace{-0.4cm}
	\label{fig: system}
	\end{center}
\end{figure}

We consider the wireless system setup illustrated in Fig.~\ref{fig: system}, comprising an FD MIMO node, a single-antenna UE, $G$ single-bounce Scattering Points (SPs), and a passive Eve. The FD node supports simultaneous communication and sensing, by receiving the UE signal and jointly exploiting it for data decoding and for estimating the positions of both the UE and the SPs. In contrast, Eve observes the same uplink waveform and aims to both intercept the communication stream and infer the surrounding scene. In order to safeguard both communication confidentiality and sensing privacy, the transmit side of the FD node concurrently emits AN designed to impair Eve's decoding and sensing capabilities.

The FD node is equipped with $N_r$- and $N_t$-element receive and transmit Uniform Linear Arrays (ULAs), respectively, while Eve employs an $N_e$-antenna array, and is located at an unknown position $\p_E\in\Omega_E\subset\mathbb R^2$, where $\Omega_E$ is a known compact uncertainty region. The UE transmits $T$ OFDM symbols over $K$ Subcarriers (SCs) with frequencies $f_k=f_c+(k-(K+1)/2)\Delta f$, with $k=1,\ldots,K$, and $s[k,t]$ denotes the complex-valued symbol transmitted on the $k$-th SC during the $t$-th OFDM symbol, satisfying $\mathbb E\{|s[k,t]|^2\}=P_u$, where $P_u$ is the UE's transmit power. At the RX side, the FD node applies a combiner $\U[k]\in\mathbb C^{N_r\times N_r}$ with constant-modulus entries $|[\U[k]]_{i,j}|=1/\sqrt{N_r}$, while its transmit side employs a unit-norm digital precoder $\v[k]\in\mathbb C^{N_t\times 1}$, with $\|\v[k]\|=1$, which processes AN's transmission.

\subsection{Channel and Signal Models}
Assuming perfect self-interference cancellation at the FD node, enabled by a hybrid analog/digital cancellation architecture, as in~\cite{alexandropoulos2017joint}, the resulting $N_r\times 1$ post-combining observation in baseband is given by:
\begin{align}
\y_F[k,t]\triangleq\U^{\rm H}[k](\h_F[k]s[k,t]+\n_F[k,t]),
\label{eq:yF_conf}
\end{align}
where $\mathbf n_F[k,t]\sim\mathcal{CN}(\mathbf 0,\sigma^2\mathbf I_{N_r})$ is the Additive White Gaussian Noise (AWGN) at the FD node's RX, and $\h_F[k]\in\Compl^{N_r\times 1}$ is the end-to-end channel between the UE and the FD RX. In contrast, Eve observes the superposition of the UE uplink signal and the FD-generated AN through the following complex $N_e\times 1$ received vector in baseband representation:
\begin{align}
\mathbf y_E[k,t]
\triangleq
\mathbf h_E[k]s[k,t]
+
\sqrt{q_k}\mathbf r_E[k]z[k,t]
+
\mathbf n_E[k,t],
\label{eq:yE_conf}
\end{align}
where $\mathbf n_E[k,t]\sim\mathcal{CN}(\mathbf 0,\sigma^2\mathbf I_{N_e})$ denotes the AWGN at Eve, $z[k,t]\sim\mathcal{CN}(0,1)$ is the AN symbol, $q_k\ge0$ is the AN power allocated to the $k$-th SC, and $\mathbf r_E[k]\triangleq\mathbf G_E[k]\mathbf v[k]$ is the precoded AN direction at Eve, with $\mathbf G_E[k]\in\mathbb C^{N_e\times N_t}$ denoting the end-to-end FD-to-Eve channel, while $\h_E[k]\in\Compl^{N_e\times 1}$ is the end-to-end UE-to-Eve channel.

Assuming far-field propagation, the UE-to-node channel, with
$i\in\{F,E\}$ denoting either the FD node or Eve, is modeled as the
superposition of one direct path $(g=0)$ and $G$ single-bounce
reflected paths $(g=1,\ldots,G)$, as follows:
\begin{align}
\mathbf h_i[k]
\triangleq
\sum_{g=0}^{G}
\alpha_{i,g} e^{-j2\pi f_k\tau_{i,g}}
\mathbf a_i(\phi_{i,g}),
\label{eq:channel_conf}
\end{align}
where $\alpha_{i,g}$ is an unknown complex-valued path gain\footnote{The complex path gains account for propagation loss, phase rotation, and reflection-dependent attenuation, and are modeled following the geometric wideband channel formulation in~\cite{shen2010wideband1,shahmansoori2018position}.},
$\tau_{i,g}$ and $\phi_{i,g}$ denote the corresponding propagation
delay and Angle of Arrival (AoA), respectively, and
$\mathbf a_i(\cdot)$ is the RX-array steering vector.

Let the FD node, UE, and $g$-th SP be located at
$\mathbf p_F$, $\mathbf p_U$, and $\mathbf p_g$, respectively, in a
two-dimensional Cartesian coordinate system. For the FD
RX, the direct-path AoA and delay are given by
\begin{align}
\phi_{F,0}
&\triangleq
\operatorname{atan2}\left([\mathbf p_U-\mathbf p_F]_2,
[\mathbf p_U-\mathbf p_F]_1\right), \nonumber\\
\tau_{F,0}
&\triangleq
\|\mathbf p_U-\mathbf p_F\|/c,
\end{align}
where $\operatorname{atan2}(\cdot,\cdot)$ denotes the four-quadrant
inverse tangent function and $c$ is the speed of light. Similarly, for the
reflected path associated with each $g$-th SP the
corresponding AoA and delay are given as:
\begin{align}\label{eq: 5}
\phi_{F,g}
&\triangleq
\operatorname{atan2}\left([\mathbf p_g-\mathbf p_F]_2,
[\mathbf p_g-\mathbf p_F]_1\right), \nonumber\\
\tau_{F,g}
&\triangleq
({\|\mathbf p_U-\mathbf p_g\|+\|\mathbf p_g-\mathbf p_F\|})/{c}.
\end{align}
The corresponding Eve-side parameters are defined analogously by
replacing $\mathbf p_F$ with $\mathbf p_E$ and expressing the AoAs with respect to
Eve's local array coordinate system. For a half-wavelength-spaced ULA,
the FD RX steering vector is $\mathbf a_F(\phi)=\frac{1}{\sqrt{N_r}}\left[1,e^{j\pi\sin\phi},\ldots,e^{j\pi(N_r-1)\sin\phi}\right]^{\rm T},$
with an analogous definition for Eve's array response. The FD-to-Eve channel $\mathbf G_E[k]$, $k=1,\ldots,K$, is modeled as a Line of Sight (LoS)-dominated downlink interference channel. Reflected FD-to-Eve components are assumed to be either negligible, unresolved, or absorbed into the effective complex gain \cite{Ayach2014SparsePrecoding}. Accordingly, $\mathbf G_E[k]$ is described using the corresponding FD TX-array and Eve RX-array responses, propagation delay, angles, and complex path gains, analogously to the geometric channel model in \eqref{eq:channel_conf}--\eqref{eq: 5}; see~\cite{Ayach2014SparsePrecoding}.

\subsection{Communication-Secrecy and Sensing-Privacy Metrics}
We quantify Eve's capabilities through two complementary metrics: the worst-case secrecy rate for UE data protection and the eavesdropping Position Error Bound (PEB) for scene-privacy protection. The former measures Eve's ability to decode the uplink stream, while the latter measures their best achievable accuracy in estimating the physical positions of the UE and all SPs in their vicinity.
\subsubsection{Sensing Privacy}
For the eavesdropping analysis, we assume the availability of a spatial region $\mathbf p_E\in\Omega_E$ for Eve's possible locations, and stack Eve's observations over all SCs for a given OFDM symbol index $t$ as $\mathbf y_E[t]\triangleq[\mathbf y_E^{\rm T}[1,t],\ldots,\mathbf y_E^{\rm T}[K,t]]^{\rm T}\in\Compl^{KN_e\times 1}$. Since the AN symbol is zero-mean and independent of both the UE signal and the thermal noise, it does not affect the mean of Eve's observation. Instead, it increases the observation covariance along the effective AN direction. Therefore, $\mathbf y_E[t]\sim\mathcal{CN}(\boldsymbol{\mu}_E,\mathbf R_E)$, where $\boldsymbol{\mu}_E\triangleq[\boldsymbol{\mu}_{E,1}^{\rm T},\ldots,\boldsymbol{\mu}_{E,K}^{\rm T}]^{\rm T}\in\Compl^{KN_e\times 1},$ with $\boldsymbol{\mu}_{E,k}\triangleq\mathbf h_E[k]s[k,t],$ and $\mathbf R_E\in\Compl^{KN_e\times KN_e}$ being block diagonal, i.e., $\mathbf R_E\triangleq\operatorname{blkdiag}\left(\mathbf R_{E,1},\ldots,\mathbf R_{E,K}\right),$ with the $k$-th covariance block expressed as follows:
\begin{align}
\mathbf R_{E,k}
\triangleq
\sigma^2\mathbf I_{N_e}
+
q_k\mathbf r_E[k]\mathbf r_E^{\rm H}[k].
\label{eq:covE_conf}
\end{align}

Let the physical scene vector be $\mathbf p_s
\triangleq
[
\mathbf p_U^{\rm T},
\mathbf p_1^{\rm T},
\ldots,
\mathbf p_G^{\rm T}
]^{\rm T}
\in\mathbb R^{2(G+1)\times 1}.$ For a fixed Eve location, define the complete unknown parameter vector as $\boldsymbol\eta_E
\triangleq[\mathbf p_s^{\rm T},\boldsymbol\beta_E^{\rm T}]^{\rm T},$ where $\boldsymbol\beta_E$ collects the real and imaginary parts of the UE-to-Eve path gains, i.e., $\boldsymbol\beta_E\triangleq[
\Re\{\alpha_{E,0}\},
\Im\{\alpha_{E,0}\},
\ldots,
\Re\{\alpha_{E,G}\},
\Im\{\alpha_{E,G}\}
]^{\rm T}.$ Since $\mathbf G_E[k]$ is modeled as a LoS-dominated channel with negligible reflected components, the effective AN direction
$\mathbf r_E[k]$ in \eqref{eq:yE_conf} is independent of the scene parameters. Hence, the $(m,n)$-th element of the FIM $\J_E\in\Compl^{4(G+1)\times 4(G+1)}$ can be derived via the Slepian--Bangs \cite{shen2010wideband1} formula, as follows:
\begin{align}
[\mathbf J_E]_{m,n}
\triangleq
2\Re\left\{
\frac{\partial \boldsymbol\mu_E^{\rm H}}
{\partial[\boldsymbol\eta_E]_m}
\mathbf R_E^{-1}
\frac{\partial \boldsymbol\mu_E}
{\partial[\boldsymbol\eta_E]_n}
\right\}
\label{eq:FIM_E_conf}
\end{align}
Partitioning $\mathbf J_E$ with respect to the position parameters $\mathbf p_s$ and the nuisance gains $\boldsymbol\beta_E$ gives
\begin{align}
\mathbf J_E
=
\begin{bmatrix}
\mathbf J_{E,pp} & \mathbf J_{E,p\beta}\\
\mathbf J_{E,\beta p} & \mathbf J_{E,\beta\beta}
\end{bmatrix},
\end{align}
where $\mathbf J_{E,pp}\in\mathbb R^{2(G+1)\times 2(G+1)}$ corresponds to the physical position parameters, $\mathbf J_{E,\beta\beta}\in\mathbb R^{2(G+1)\times 2(G+1)}$ corresponds to the real-valued nuisance gain vector, and the off-diagonal blocks capture their coupling \cite{shen2010wideband1}. Eliminating the nuisance gains through the Schur complement yields the eavesdropping Equivalent FIM (EFIM) for the physical scene positions:
\begin{align}
\mathbf J_{E,p}
=
\mathbf J_{E,pp}
-
\mathbf J_{E,p\beta}
\mathbf J_{E,\beta\beta}^{-1}
\mathbf J_{E,\beta p}.
\label{eq:efimE_pos_conf}
\end{align}
Accordingly, the position-domain CRB matrix is given by $\mathbf C_E \triangleq\mathbf J_{E,p}^{-1}.$ The aggregate eavesdropping PEB is then defined as $\mathrm{PEB}\triangleq\sqrt{\mathrm{Tr}\left\{\mathbf C_E\right\}}$.

Given the uncertainty region $\Omega_E$, sensing privacy is quantified by the worst-case eavesdropping estimation accuracy over all possible Eve locations. Since a larger CRB corresponds to poorer scene estimation capability at Eve, the relevant worst case is the minimum eavesdropping CRB over $\Omega_E$, defined as
\begin{align}
C_E^{\rm wc}
\triangleq
\inf_{\mathbf p_E\in\Omega_E}
\mathrm{CRB}_E(\mathbf p_E).
\label{eq:gap_conf}
\end{align}

\subsubsection{Communication Secrecy}
We next quantify the ability of the FD node to protect UE data from Eve. Since the FD RX is assumed to perfectly cancel its self-interference, the legitimate link is affected only by thermal noise, whereas Eve observes the UE signal in the presence of FD-generated AN. The legitimate post-combining Signal-to-Interference-plus-Noise Ratio (SINR) at the FD is given by:
\begin{align}
\gamma_F
\triangleq
P_u
\sum_{k=1}^{K}
\mathbf h_F^{\rm H}[k]\mathbf P_U[k]\mathbf h_F[k],
\end{align}
where $P_U[k]\triangleq \sigma^{-2}\mathbf U[k]\mathbf U^{\rm H}[k]$. At Eve, the AN contributes to the interference-plus-noise covariance in \eqref{eq:covE_conf}. Therefore, for a given Eve location \(\mathbf p_E\), the eavesdropping SINR is:
\begin{align}
\gamma_E
\triangleq
P_u\sum_{k=1}^{K}
\mathbf h_E^{\rm H}[k]
\mathbf R_{E,k}^{-1}
\mathbf h_E[k].
\label{eq:gammaE_conf}
\end{align}
This expression showcases that the transmitted AN affects Eve only through the inverse covariance matrix: increasing \(q_k\), $\forall k$ reduces the contribution of the UE signal along the effective AN direction observed by Eve. Since Eve's exact position is unknown, communication secrecy is evaluated in the worst case over the uncertainty region \(\Omega_E\), i.e.,
\begin{align}
R_c^{\rm wc}
\triangleq
\inf_{\mathbf p_E\in\Omega_E}
\left[
\log_2(1+\gamma_F)
-
\log_2\!\left(1+\gamma_E\right)
\right]^+.
\label{eq:Rwc_conf}
\end{align}

\section{Proposed FD-Enabled Secure ISAC}\label{sec: 3}
This section characterizes the finite-power limits of AN-aided protection against malicious eavesdropping. We first derive the irreducible eavesdropping SINR floor and the asymptotic CRB level induced by the inverse interference-plus-noise covariance. These limits are then used to establish local and robust secrecy--privacy feasibility conditions, and to develop a branch-and-bound power-allocation method.

\subsection{Irreducible Eavesdropping Floors}
For a fixed Eve location $\mathbf p_E$, the eavesdropping SINR in
\eqref{eq:gammaE_conf} depends on the inverse of the interference-plus-noise
covariance matrix $\R_{E}$. Since this covariance is a rank-one perturbation of the white-noise covariance, applying the Sherman--Morrison identity \cite{boyd2004convex} to each block of $\R_E$ yields
\begin{align}\nonumber
\mathbf R_{E,k}^{-1}
=
\sigma^{-2}\mathbf I_{N_e}
-
\frac{q_k}{\sigma^4\left(1+c_k(\mathbf p_E)q_k\right)}
\mathbf r_E[k;\mathbf p_E]\mathbf r_E^{\rm H}[k;\mathbf p_E],
\end{align}
where
$c_k(\mathbf p_E)\triangleq\sigma^{-2}\|\mathbf r_E[k;\mathbf p_E]\|^2$.
Substituting this expression into \eqref{eq:gammaE_conf} yields the following expression:
\begin{align}
\gamma_E(\mathbf p_E;\mathbf q)
=
\sum_{k=1}^{K}
\left(
a_k(\mathbf p_E)
-
\frac{b_k(\mathbf p_E)q_k}
{1+c_k(\mathbf p_E)q_k}
\right),
\label{eq:sinr_sm_conf}
\end{align}
where $a_k(\mathbf p_E)\triangleq\frac{P_u}{\sigma^2}\|\mathbf h_E[k;\mathbf p_E]\|^2$ and $b_k(\mathbf p_E)\triangleq\frac{P_u}{\sigma^4}\left|\mathbf r_E^{\rm H}[k;\mathbf p_E]\mathbf h_E[k;\mathbf p_E]\right|^2$.
Here, $a_k(\mathbf p_E)$ is the eavesdropping contribution on the
$k$-th SC in the absence of AN, while the second term quantifies the
SINR reduction produced by AN. This reduction depends on the alignment
between Eve's useful channel $\mathbf h_E[k;\mathbf p_E]$ and the
effective AN direction $\mathbf r_E[k;\mathbf p_E]$. Therefore, AN can
suppress only the component of Eve's channel that lies in the
AN-illuminated subspace. As $q_k\rightarrow\infty$, this reduction
saturates at $b_k(\mathbf p_E)/c_k(\mathbf p_E)$, yielding the
irreducible eavesdropping SINR floor:
\begin{align}
\gamma_E^{\infty}(\mathbf p_E)
=
\sum_{k=1}^{K}
\left(
a_k(\mathbf p_E)
-
\frac{b_k(\mathbf p_E)}
{c_k(\mathbf p_E)}
\right),
\label{eq:floorC_conf}
\end{align}
assuming that all SCs satisfy $c_k(\mathbf p_E)>0$.

The same inverse-covariance structure determines the effect of AN on
Eve's sensing capability. Thus, AN reduces the inverse covariance only along the effective AN
direction $\mathbf r_E[k;\mathbf p_E]$. Since
$q_k/(1+c_k(\mathbf p_E)q_k)$ is continuous and nondecreasing in
$q_k$, increasing the AN power decreases
$\mathbf R_{E,k}^{-1}$ in the Loewner order \cite{kay1993fundamentals}. Hence, Eve's FIM is
nonincreasing in $\mathbf q\triangleq[q_1,\ldots,q_K]\in\mathbb{R}^{K\times 1}$. Consequently, the position-domain EFIM
$\mathbf J_{E,p}(\mathbf p_E;\mathbf q)$ is nonincreasing, while the
corresponding position-domain CRB:
\begin{align}
\mathrm{CRB}_E(\mathbf p_E;\mathbf q)
\triangleq
\operatorname{tr}\left\{
\mathbf J_{E,p}^{-1}(\mathbf p_E;\mathbf q)
\right\}
\label{eq:crbE_pair_conf}
\end{align}
is continuous and component-wise nondecreasing in $\mathbf q$.

The zero-AN and asymptotic eavesdropping CRB levels are defined,
respectively, as
\begin{align}
C_E^0(\mathbf p_E)
&\triangleq
\mathrm{CRB}_E(\mathbf p_E;\mathbf 0_{K\times 1}),
\\
C_E^{\infty}(\mathbf p_E)
&\triangleq
\operatorname{tr}\left\{
\left(\mathbf J_{E,p}^{\infty}(\mathbf p_E)\right)^{-1}
\right\}.
\label{eq:CEinf_conf}
\end{align}
Here, $\mathbf J_{E,p}^{\infty}(\mathbf p_E)$ is the limiting
position-domain EFIM obtained as $q_k\rightarrow\infty$ $\forall k$ active
SCs. Equivalently, it is computed by replacing each inverse covariance
block in the FIM with:
\begin{align}
\left(\R_{E,k}^{\infty}(\mathbf p_E)\right)^{-1}
=
\frac{1}{\sigma^2}
\left(
\mathbf I_{N_e}
-
\frac{\mathbf r_E[k;\mathbf p_E]\mathbf r_E^{\rm H}[k;\mathbf p_E]}
{\|\mathbf r_E[k;\mathbf p_E]\|^2}
\right).
\label{eq:REinv_inf_conf}
\end{align}
This limiting inverse covariance projects Eve's observation onto the
subspace orthogonal to the AN direction. If the resulting
$\mathbf J_{E,p}^{\infty}(\mathbf p_E)$ is singular, then, Eve loses
identifiability of at least one protected position direction in the
infinite-AN limit, and
$C_E^{\infty}(\mathbf p_E)\rightarrow\infty$.

\subsection{Local and Robust Feasibility}
In this section we state the main analytical contribution of the paper: local and robust conditions under which AN can jointly guarantee communication secrecy and sensing privacy. The resulting thresholds reveal when finite AN power is sufficient, when no AN injection is required, and when the target secrecy--privacy pair is fundamentally infeasible.
\begin{proposition}
Let $R_0$ denote the target secrecy rate and let $\Gamma_0$ denote the minimum admissible eavesdropping CRB. For a fixed Eve location $\mathbf p_E$, define $\zeta\triangleq(1+\gamma_F)2^{-R_0}-1$ \cite{10143983}. The communication-secrecy constraint then falls into the following regimes:
    \begin{enumerate}
        \item If $\zeta\ge\sum_{k=1}^{K}a_k(\mathbf p_E)$, then,
        $R_c(\mathbf p_E;\mathbf q)\ge R_0$ holds at $\mathbf q=\mathbf 0_{K\times 1}$.
        \item If
        $\gamma_E^{\infty}(\mathbf p_E)<\zeta<
        \sum_{k=1}^{K}a_k(\mathbf p_E)$, then, finite AN power can satisfy
        $R_c(\mathbf p_E;\mathbf q)\ge R_0$.
        \item If $\zeta\le\gamma_E^{\infty}(\mathbf p_E)$, then, the
        secrecy target is infeasible with finite AN power.
    \end{enumerate}
Similarly, the sensing-privacy constraint is characterized by the following regimes:
    \begin{enumerate}
        \item If $\Gamma_0\le C_E^0(\mathbf p_E)$, then,
        $\mathrm{CRB}_E(\mathbf p_E;\mathbf q)\ge\Gamma_0$
        holds at $\mathbf q=\mathbf 0_{K\times 1}$.
        \item If
        $C_E^0(\mathbf p_E)<\Gamma_0<C_E^{\infty}(\mathbf p_E)$, then,
        finite AN power can satisfy
        $\mathrm{CRB}_E(\mathbf p_E;\mathbf q)\ge\Gamma_0$.
        \item If $\Gamma_0\ge C_E^{\infty}(\mathbf p_E)$, then, the
        sensing-privacy target is infeasible with finite AN power.
    \end{enumerate}

\end{proposition}

\begin{IEEEproof}
Equation~\eqref{eq:sinr_sm_conf} expresses Eve's SINR as the zero-AN value minus an AN-induced reduction term on each $k$-th SC. Specifically, the reduction on the $k$-th SC is
$b_k q_k/(1+c_k q_k)$, which is continuous and monotonically increasing in $q_k$. Its maximum attainable value is $b_k/c_k$, obtained in the limit $q_k\rightarrow\infty$. Therefore, Eve's SINR can be decreased from its zero-AN value $\sum_{k=1}^K a_k$ only to the irreducible floor $\gamma_E^{\infty}$. If the required threshold $\zeta$ is above the zero-AN SINR, no AN is needed; if it lies strictly between the zero-AN SINR and the floor, continuity guarantees that a finite AN power achieves it; and if it is below the floor, no finite AN power can satisfy the secrecy constraint.

For the sensing constraint, \eqref{eq:crbE_pair_conf} shows that increasing the AN power decreases the inverse covariance $\mathbf R_E^{-1}$ in the Loewner order. Since the adopted FD-to-Eve AN channel is independent of the protected scene parameters, the covariance-derivative term in the FIM is absent. Thus, Eve's FIM depends on AN only through this inverse-covariance weighting, and it is nonincreasing with $\mathbf q$. After nuisance-gain elimination, the same ordering holds for the position-domain EFIM $\mathbf J_{E,p}$. Consequently, matrix inversion reverses the ordering, and the $\mathrm{CRB}_E(\mathbf p_E;\mathbf q)$ is continuous and componentwise nondecreasing in the AN powers: its smallest value is the zero-AN value $C_E^0(\mathbf p_E)$, and its largest attainable finite-power limit is $C_E^{\infty}(\mathbf p_E)$. Hence, the sensing-privacy target is already satisfied if it is below $C_E^0(\mathbf p_E)$, is achievable with finite AN power if it lies strictly between $C_E^0(\mathbf p_E)$ and $C_E^{\infty}(\mathbf p_E)$, and is infeasible if it exceeds the asymptotic limit.
\end{IEEEproof}

\begin{proposition}
For fixed RX combiners and AN directions, finite-power robust
pair feasibility over $\Omega_E$ requires
\begin{align}
R_0
&<
\log_2(1+\gamma_F)
-
\log_2\left(
1+
\sup_{\mathbf p_E\in\Omega_E}
\gamma_E^{\infty}(\mathbf p_E)
\right),
\label{eq:conditionR_conf}\\
\Gamma_0
&<
\inf_{\mathbf p_E\in\Omega_E}
C_E^{\infty}(\mathbf p_E).
\label{eq:conditionS_conf}
\end{align}
If the channel responses, derivative matrices, and AN directions are
continuous over the compact region $\Omega_E$, then
\eqref{eq:conditionR_conf}--\eqref{eq:conditionS_conf} are also
sufficient, i.e., a feasible finite-AN solution exists.
\end{proposition}

\begin{IEEEproof}
The necessity of \eqref{eq:conditionR_conf} follows by applying the local secrecy threshold uniformly over Eve's position uncertainty region. For any fixed $\mathbf p_E\in\Omega_E$, even unbounded AN power can reduce Eve's SINR only down to $\gamma_E^{\infty}(\mathbf p_E)$. Hence, if the secrecy constraint is to hold for every possible Eve location, the target rate $R_0$ must be smaller than the secrecy rate obtained under the largest irreducible SINR floor over $\Omega_E$, which yields \eqref{eq:conditionR_conf}. Similarly, for sensing privacy, the largest Eve CRB attainable at location $\mathbf p_E$ is $C_E^{\infty}(\mathbf p_E)$. Hence, to guarantee $\mathrm{CRB}_E(\mathbf p_E;\mathbf q)\ge\Gamma_0$ $\forall\mathbf p_E\in\Omega_E$ with finite AN power, the threshold $\Gamma_0$ must be strictly smaller than the smallest asymptotic CRB level over the uncertainty region, yielding condition~\eqref{eq:conditionS_conf}.

To prove sufficiency, consider the common power allocation $\mathbf q=t\mathbf 1_K$ with $t\ge0$. For each fixed $\mathbf p_E$, the eavesdropping SINR $\gamma_E(\mathbf p_E;t\mathbf 1_{K\times 1})$ is continuous and decreases to $\gamma_E^{\infty}(\mathbf p_E)$ as $t\rightarrow\infty$. Likewise, $\mathrm{CRB}_E(\mathbf p_E;t\mathbf 1_{K\times 1})$ is continuous and increases to $C_E^{\infty}(\mathbf p_E)$. Since the channel responses, derivative matrices, and AN directions are continuous over the compact set $\Omega_E$, these pointwise convergences are uniform over $\Omega_E$. Therefore, if \eqref{eq:conditionR_conf}--\eqref{eq:conditionS_conf} hold, there exists a finite $t$ such that the secrecy rate constraint and the CRB threshold are simultaneously satisfied for every $\mathbf p_E\in\Omega_E$.
\end{IEEEproof}

\begin{remark}
The feasibility conditions reveal a common subspace limitation. AN can
suppress Eve's communication SINR only when the component of
$\mathbf h_E[k]$ is aligned with the effective AN direction
$\mathbf r_E[k]$. Similarly, it can degrade Eve's sensing accuracy only
through the sensing-derivative components that are observed along the
AN-illuminated subspace. In the high-AN regime, Eve's inverse covariance converges to a projection onto the subspace orthogonal to $\mathbf r_E[k]$; hence, any communication or sensing information contained in that orthogonal subspace remains unaffected, and further AN power cannot improve secrecy or sensing privacy.
\end{remark}

\subsection{Power Allocation Optimization}
We now turn the feasibility analysis into a robust AN power-allocation problem under uncertain Eve location. Specifically, we approximate the compact region $\Omega_E$ by a finite set of representative points $\{\mathbf p_\ell\}_{\ell=1}^{L}$. For fixed RX combiners and AN directions, the proposed robust AN power-allocation problem is formulated as follows:
\begin{align}
\min_{\mathbf q\succeq\mathbf 0}\quad 
& \mathbf 1_K^{\rm T}\mathbf q
\label{eq:opt_conf}\\
\text{s.t.}\quad 
& R_c(\mathbf p_\ell;\mathbf q)\ge R_0, \quad \ell=1,\ldots,L, \nonumber\\
& \mathrm{CRB}_E(\mathbf p_\ell;\mathbf q)\ge \Gamma_0, 
\quad \ell=1,\ldots,L, \nonumber
\end{align}
Under the adopted channel models, both functions in the constraints are monotone in $\mathbf q$: Eve's SINR is componentwise nonincreasing in the AN powers, whereas Eve's CRB $\mathrm{CRB}_E(\mathbf p_\ell;\mathbf q)$ is componentwise nondecreasing. Hence, the feasible set of \eqref{eq:opt_conf} is upward closed, i.e., if $\mathbf q$ is feasible, then any $\mathbf q'\succeq\mathbf q$ is also feasible \cite{boyd2004convex}.

This monotonicity enables the application of a certifiable branch-and-bound procedure, as in~\cite{boyd2004convex}. Given any feasible upper point $\bar{\mathbf q}\in\mathbb R_+^{K\times 1}$, the search is restricted to $[\mathbf 0,\bar{\mathbf q}]$. For each box $\mathcal B=[\mathbf l,\mathbf u]$, monotonicity provides two pruning rules: if $\mathbf u\in\mathbb R_+^{K\times 1}$ violates any sampled secrecy or CRB constraint, then $\mathcal B$ is infeasible; if $\mathbf l\in\mathbb R_+^{K\times 1}$ is feasible, then it is the minimum-power feasible point in $\mathcal B$. The algorithm repeatedly partitions the active box with the smallest lower bound $\mathbf 1_K^{\rm T}\mathbf l$ until the gap to the best feasible upper bound is below $\varepsilon$, yielding an $\varepsilon$-globally optimal solution for the sampled problem. The initial $\bar{\mathbf q}$ can be obtained by increasing a common power $q\mathbf 1_K$ until all sampled constraints are met, and the continuous-region accuracy depends on the sampling density over $\Omega_E$. Algorithm~\ref{alg:bb_power} summarizes the procedure.

\begin{algorithm}[t]
\caption{Branch-and-Bound for AN Power Allocation}
\label{alg:bb_power}
\begin{algorithmic}[1]
\REQUIRE $\{\mathbf p_\ell\}_{\ell=1}^{L}$, $\bar{\mathbf q}$, $\varepsilon>0$
\STATE Initialize active box set $\mathcal A\gets\{[\mathbf 0,\bar{\mathbf q}]\}$.
\STATE Set incumbent $\mathbf q_{\rm inc}\gets\bar{\mathbf q}$ and $Z_{\rm UB}\gets\mathbf 1_K^{\rm T}\bar{\mathbf q}$.
\WHILE{$\mathcal A\neq\emptyset$}
    \STATE Set $Z_{\rm LB}\gets\min_{[\mathbf l,\mathbf u]\in\mathcal A}\mathbf 1_K^{\rm T}\mathbf l$.
    \IF{$Z_{\rm UB}-Z_{\rm LB}\le\varepsilon$}
        \STATE \textbf{break}
    \ENDIF
    \STATE Select $\mathcal B=[\mathbf l,\mathbf u]\in\mathcal A$ with the smallest $\mathbf 1_K^{\rm T}\mathbf l$.
    \STATE Remove $\mathcal B$ from $\mathcal A$.
    \IF{$\mathbf u$ violates the secrecy or CRB constraints}
        \STATE Discard $\mathcal B$.
    \ELSIF{$\mathbf l$ satisfies all secrecy and CRB constraints}
        \IF{$\mathbf 1_K^{\rm T}\mathbf l<Z_{\rm UB}$}
            \STATE Set $\mathbf q_{\rm inc}\gets\mathbf l$ and $Z_{\rm UB}\gets\mathbf 1_K^{\rm T}\mathbf l$.
        \ENDIF
    \ELSE
        \STATE Choose $j\gets\arg\max_i\{[\mathbf u]_i-[\mathbf l]_i\}$ and $m\gets([\mathbf l]_j+[\mathbf u]_j)/2$. 
        \STATE Form $\mathbf u^{(1)}\gets\mathbf u$ with $[\mathbf u^{(1)}]_j=m$. Form $\mathbf l^{(2)}\gets\mathbf l$ with $[\mathbf l^{(2)}]_j=m$.
        \STATE Add $[\mathbf l,\mathbf u^{(1)}]$ and $[\mathbf l^{(2)},\mathbf u]$ to $\mathcal A$.
    \ENDIF
\ENDWHILE
\RETURN $\mathbf q_{\rm inc}$
\end{algorithmic}
\end{algorithm}

\section{Numerical Results and Discussion}
This section validates the proposed FD-enabled ISAC safeguarding framework through Monte Carlo simulations. We focus on the power cost of joint safeguarding by computing the minimum AN power required to satisfy prescribed secrecy--privacy target pairs over Eve's position uncertainty region. 

\subsection{Simulation Setup}
A two-dimensional deployment having the main simulation parameters summarized in Table~\ref{tab:sim_params} has been considered. The FD MIMO node was assumed located at the origin, the UE is located at $\mathbf p_U=[20,10]^{\rm T}$, and the two SPs were placed at $\mathbf p_1=[15,22]^{\rm T}$ and $\mathbf p_2=[32,6]^{\rm T}$. Eve's nominal position was selected as $\mathbf p_{E}=[52,22]^{\rm T}$ and was assumed to lie within a circular uncertainty region centered at $\mathbf p_{E}$ with radius $5$ meters. The noise variance was computed as $\sigma^2=N_0K\Delta f N_F$, where $N_0$ is the noise power spectral density and $N_F$ is the RX noise figure. The analog RX combiner was set following the FD RX-array steering vectors to match the dominant UE-to-FD sensing and communication directions, and was kept fixed during AN power allocation. Similarly, the digital AN precoder was selected from the FD TX-array steering vectors as a unit-norm beam steered toward Eve's uncertainty region. All numerical results were obtained via averaging over $500$ Monte Carlo simulations.

\begin{table}[!t]
\centering
\caption{Simulation Parameters.}
\label{tab:sim_params}
\begin{tabular}{|c|c|c|c|}
\hline
\textbf{Parameter} & \textbf{Value} & \textbf{Parameter} & \textbf{Value} \\
\hline\hline
$f_c$ & $20$ GHz & $\Delta f$ & $120$ kHz \\
\hline
$K$ & $128$ & $P_u$ & $0$ dBm \\
\hline
$N_r$, $N_t$ & $16$, $16$ & $N_e$ & $8$ \\
\hline
$G$ & $2$  & $N_F$ & $8$ dB \\
\hline
$N_0$ & $-174$ dBm/Hz & $L$, $\epsilon$ & $15$, $10^{-3}$ \\
\hline
\end{tabular}
\vspace{-0.4cm}
\end{table}

\subsection{Secrecy-Privacy AN Power Tradeoff}

\begin{figure}[!t]
	\begin{center}
	\includegraphics[width=0.97\columnwidth]{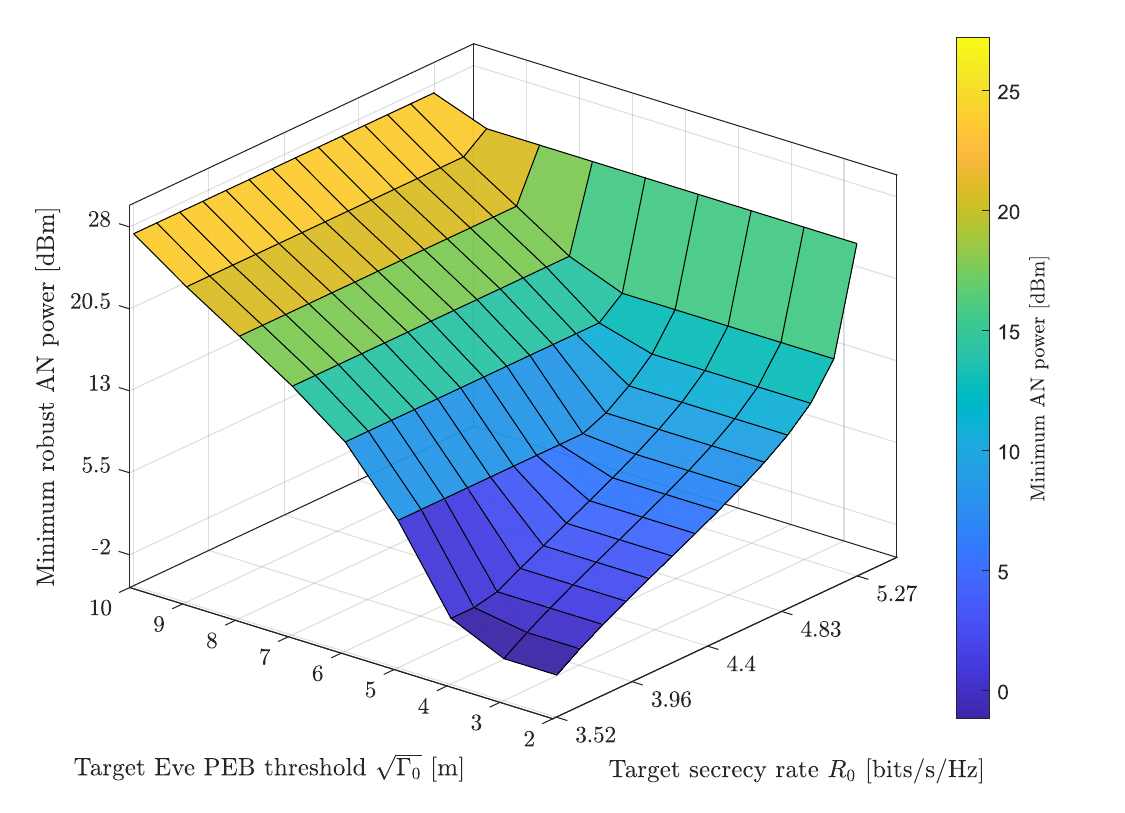}
	\caption{\small{Minimum robust AN power required to jointly satisfy the target secrecy rate and the sensing-privacy $(R_0,\sqrt{\Gamma_0})$ thresholds. The surface showcases the coupled secrecy--privacy power cost: increasing either the communication-secrecy requirement or the minimum admissible PEB at Eve requires more AN power, with the steep growth near the high-target region indicating operation close to the finite-power feasibility boundary.
   }}
    \vspace{-0.4cm}
	\label{fig: tradeoff}
	\end{center}
\end{figure}

Figure~\ref{fig: tradeoff} showcases the minimum AN power required to satisfy a target secrecy--privacy threshold pair \((R_0,\sqrt{\Gamma_0})\). The horizontal axes represent the target secrecy rate $R_0$ and the target PEB threshold $\sqrt{\Gamma_0}$ at Eve, while the vertical axis and color scale report the minimum total AN power obtained by the branch-and-bound solver. Hence, each point on the surface gives the least AN power needed to jointly guarantee \(R_c^{\rm wc}\ge R_0\) and \(C_E^{\rm wc}\ge\Gamma_0\) over the sampled Eve uncertainty region. As shown, the surface increases along both dimensions, confirming that communication secrecy and sensing privacy compete for the same AN resource. The steep rise in the high rate and PEB region indicates that the requested pair approaches the finite-power feasibility boundary of Section~\ref{sec: 3}. This is consistent with the subspace-limited analysis: once Eve's remaining information lies largely outside the AN-illuminated subspace, increasingly large AN power is required for small additional gains. Equivalently, the surface can be interpreted as a design map that identifies which secrecy--privacy target pairs are supportable under a given AN power budget.

\section{Conclusion}
This paper proposed an FD-enabled ISAC safeguarding framework against a passive multi-antenna Eve that targets both data interception and scene inference. We formulated joint secrecy--privacy protection through worst-case secrecy rate and eavesdropping CRB constraints over an uncertain Eve location region. By deriving the relevant eavesdropping information measures and exploiting their dependence on the inverse interference-plus-noise covariance, we characterized finite-power local and robust feasibility conditions. The presented analysis revealed a common subspace limitation: AN improves protection only along Eve's communication and sensing-derivative directions illuminated by the effective AN subspace. We further formulated a monotone branch-and-bound AN power allocation method for the sampled robust problem. Numerical results quantified the AN power required to satisfy joint secrecy--privacy constraints and confirmed the finite-power tradeoff predicted by the analysis.

\bibliographystyle{IEEEtran}
\bibliography{ms2}
\end{document}